\documentstyle[aps,preprint,eqsecnum,pra]{revtex}

\begin{document}

\draft

\tighten

\title{Interference of Bose condensates}
\author{M.~Naraschewski,\cite{LMU} H.~Wallis, and
A.~Schenzle\cite{LMU}}
\address{Max--Planck--Institut f\"ur Quantenoptik\\
Hans--Kopfermann--Stra{\ss}e 1, D--85748~Garching, Germany}
\author{J. I. Cirac}
\address{Departamento de Fisica Aplicada, Universidad de Castilla-La
Mancha,\\
13071 Ciudad Real, Spain}
\author{P. Zoller}
\address{Institut f\"ur theoretische Physik, Universit\"at
Innsbruck, Technikerstr. 25, Austria}

\date{January 31, 1996; revised April 25, 1996}

\maketitle

\begin{abstract}
We investigate the prospects of atomic interference using samples of
Bose condensed atoms. First we show the ability of two independent
Bose condensates to create an interference pattern. This holds even if
both condensates are described by Fock states. Thus, the existence of
an experimental signature for a broken gauge symmetry, seen in a
single run of the experiment, is not necessarily reflected by a broken
symmetry on the level of the quantum mechanical state vector. Based on
these results, we simulate numerically a recent experiment with two
independent Bose condensates [K.B.~Davis et al., PRL 75, 3969
(1995)]. The existence of interference fringes is predicted based on
the nonlinear Schr\"odinger equation. Finally we study theoretically
the influence of finite temperatures on the visibility of the
interference in a double pinhole configuration.
\end{abstract}


\narrowtext

\section{Introduction}
\label{Introduction}
Recent realizations of Bose-Einstein condensation (BEC) in dilute and
ultracold gases \cite{anderson95,bradley95,ketterle95} have attracted
vivid interest. It is hoped that the study of those experimental
systems will give new insight into the physics of BEC. Since the
current understanding of BEC is largely influenced by the concept of
a macroscopic wave function, the study of this feature is of foremost
importance. The investigation of interference phenomena should be
perfectly apted for this purpose. Another motivation for the study of
interference properties is the envisioned development of a new source
of atoms, based on BEC, with high flux and coherence, that
is expected to stimulate atomic interference experiments.

Although interference is a well known phenomenon in quantum
statistically degenerate systems (cf. Josephson effect), surprisingly
little quantitative analysis has been performed up to now, to
investigate the interference properties of a Bose condensate. In this
paper we study theoretically two interference experiments that could
be performed with the technology of current BEC experiments
\cite{anderson95,bradley95,ketterle95}.

In Sec.\ \ref{twocond} we focus on the general possibility of
interference between two independent Bose condensates. This topic is
closely related to earlier studies, showing interference between two
independent laser beams \cite{mandel67,ou88}. Interference of atoms,
originating from independent sources, differs qualitatively from usual
interference experiments, where atoms interfere only with themselves.
In this case the phase of the interference pattern is uniquely defined
by the geometry of the setup. For two independent condensates,
however, an interference pattern will be exhibited with a phase that
unpredictably varies between {\em different runs} of the experiment.
This implies that the system possesses some symmetry that is
spontaneously broken in a single run of the experiment. An intuitive
explanation for this interference follows from the common notion of
spontaneously broken gauge symmetry, implying the existence of a
macroscopic wave function. We show, however, that the system behaves
identically under circumstances where the broken symmetry is not
reflected by the quantum mechanical state vector, e.g.\ if the two
condensates are initially in Fock states. In this case, it is still
possible to derive from quantum mechanics the existence of
interference patterns in a single run.

Having shown the interference of two independent Bose condensates in
quite general terms, we turn more realistic in Sec.\ \ref{numsim}.
This includes accounting for the finite size of experimental
condensates and the important influence of atom interactions. Finite
temperature effects, however, are still neglected. In principle, a
setup for interference of two condensates has been realized in the BEC
experiment of \cite{ketterle95}. There, two condensates are stored in
a magnetic trap, separated by a light beam. After releasing them from
the trap, they expand and eventually overlap. However, no
interference has been observed in the reported experiment
\cite{ketterle95}. We simulate this experiment numerically using the
nonlinear Schr\"odinger equation. We are able to reproduce the
observed final size of the atomic cloud in good agreement with
\cite{ketterle95}. In
addition it is shown that interference fringes should exist with a
spatial period smaller than the present experimental resolution.

In Sec.\ \ref{visib} we investigate the more conventional case where a
single condensate is released from a harmonic trap and subsequently
interferes with itself in a simple double pinhole experiment. In this
calculation we allow for finite temperatures but neglect atom
interactions. It is investigated, to what extent the existence of
interference fringes might serve as a signature of Bose condensation.
It turns out that the appearance of a condensate is reflected by an
extremely sharp change in the visibility of the interference pattern
at the critical temperature.

After completion of our work we have become aware of a recent paper by
J.~Javanainen and S.~Yoo \cite{javanainen95}. These authors have studied
the interference of two
independent Bose condensates with conclusions, similar to the ones
reached in Sec.\ \ref{twocond}, but based on a complementary derivation.

\section{Interference of two independent Bose condensates}
\label{twocond}
\subsection{Spontaneously broken gauge symmetry}
\label{gauge}
In usual interference experiments (cf.\ Sec.\ \ref{visib}), atomic
beams are split apart in a suitable way, e.g.\ by a Young double slit,
and recombined afterwards thereby leading to an interference signal.
Since all atoms come from the same source and are not tracked on
their way through the interferometer one usually notes that atoms
interfere with themselves in such a situation. Here, we address the
question whether an interference signal can be exhibited as well, when
two {\em independent} Bose condensates merge.

Although, in this case atoms do not interfere with themselves but
rather with other indistinguishable atoms, it should indeed be
possible to observe such an interference under suitable conditions. A
simple understanding of this phenomenon can be achieved using the
notion of spontaneously broken gauge symmetry, that is widely regarded
as a characteristic feature of BEC. We will shortly summarize this
concept in the following.

A spontaneously broken symmetry, in general, implies that the
behaviour of a {\em single} many-particle system differs from its {\em
ensemble} average. The common paradigm for spontaneous symmetry
breaking is a ferromagnet, where a single domain may show a
magnetization while the mean magnetization, averaged over different
domains, vanishes. It is a common approach in statistical physics
\cite{huang87} to investigate the existence of such a broken symmetry
by adding a suitable small symmetry breaking field (Bogoliubov
auxiliary field) to the Hamiltonian of the system. The ground state of
this new Hamiltonian is meant to describe a particular single system
at zero temperature with broken symmetry instead of the symmetric
ensemble. The density matrix, describing the full ensemble, is then
assumed to be a symmetry preserving, incoherent superposition of
degenerate ground states, corresponding to different orientations of
the Bogoliubov field.

So as to account for several observed interference phenomena like the
Josephson effect, it is a common notion to postulate a spontaneously
broken gauge symmetry \cite{fetter71} in the case of BEC. The
corresponding Bogoliubov field consists in adding
\begin{equation}
\label{bogfield}
\epsilon\Psi(x)+\epsilon^*\Psi^\dagger(x)
\end{equation}
to the Hamiltonian, letting $\epsilon\to 0$
after having performed the thermodynamic limit.
It does not correspond to any physical interaction, in
contrast to the case of ferromagnetism. Its consequence is that a {\em
single} Bose condensate is described by a {\em coherent state} with a
distinguished phase. All measurable quantities of this particular
single condensate are then fully characterized by the complex scalar
field $\psi(x)$ with
\begin{equation}
\label{psisys}
\left<\Psi(x)\right>_S=\psi(x).
\end{equation}
Note, that the angular brackets in Eq.\ (\ref{psisys}) refer only to
the {\em subensemble} that is defined by the particular coherent
state. The field $\psi(x)$ is usually called the {\em macroscopic wave
function} of the condensate. The macroscopic wave function $\psi'(x)$
of any other condensate, seen in a different run of the same
experimental setup, can be achieved by applying the gauge
transformation
\begin{equation}
\label{symmtrans}
\Psi'(x)=\exp(i\alpha \hat N)\Psi(x)\exp(-i\alpha \hat N)
\end{equation}
with $\hat N$ being the number operator. This yields
\begin{equation}
\psi'(x) = e^{-i\alpha}\psi(x),
\end{equation}
motivating the notion of a broken gauge symmetry. In contrast to the
single system, the {\em full ensemble} does not show any phase
dependence
\begin{equation}
\label{psien}
\left<\Psi(x)\right>=0,
\end{equation}
i.e.\ it is invariant under the symmetry transformation Eq.\
(\ref{symmtrans}). The main advantage of this concept is that
quantities, measured in a {\em single run} of the experiment, can
still be expressed by simple quantum mechanical expectation values,
while the true {\em ensemble} expectation values fail to characterize
a single system properly.

So far, the analogy between spontaneous symmetry breaking in a
ferromagnet and in BEC seems to hold perfectly. However, spontaneously
broken gauge symmetry has to be taken with some caveat, as will be
indicated in the following. In the case of ferromagnetism, the ground
state of the unmodified Hamiltonian is {\em degenerate}. The effect of
the Bogoliubov field consists only in selecting a particular one of
these degenerate states for the characterization of a given domain.

Such a correspondence exists equally in the usual treatment of BEC.
Here, the density matrix of the condensate at $T=0$ is expressed in
terms of {\em coherent states}
\begin{equation}
\label{rhoBECcoh}
\rho_0=\frac{1}{\pi}\int
d^2\alpha\,\delta(|\alpha|^2-N)\,\left|\alpha\right>
\left<\alpha\right|,
\end{equation}
from which the Bogoliubov field selects a particular
$\left|\bar\alpha\right>$, due to the assumed broken gauge symmetry.

The justification of this treatment may be questioned, since the
Bogoliubov field Eq.\ (\ref{bogfield}) does not correspond to a
real physical field. In addition, atomic coherent states are no
eigenstates of the particle number and violate atom number conservation,
that is implied by fundamental
superselection rules. It is therefore not obvious why a decomposition
in terms of coherent states rather than in terms of number states is
appropriate for the characterization of a single run of the
experiment. Another subtle feature arises from the fact that the
assumed degeneracy of the ground state requires the use of the
grandcanonical ensemble with fluctuations of the particle number. In a
microcanonical ensemble with a fixed particle number $N$, however, the
ground state of the unmodified Hamiltonian may be {\em nondegenerate},
i.e.\ given by a single Fock state
\begin{equation}
\label{rhoBEC}
\rho_0=\left|N\right>\left<N\right|.
\end{equation}
Obviously, the concept of the Bogoliubov field, merely selecting one of
several existing ground states, becomes meaningless in
this case. This raises the important question whether a microcanonical
description will also show BEC with all its characteristic features.
A positive answer to this question would imply that the degeneracy of
the ground state is not essential for the understanding of BEC, in
contrast to the case of ferromagnetism. Indeed, a numerical simulation
of the quantum kinetic equation \cite{jaksch96} has shown that a
macroscopic population of a nondegenerate ground state occurs in a
system with a fixed particle number. It remains to be seen,
whether the density matrix Eq.\ (\ref{rhoBEC}) is able to reproduce
even those effects, that are commonly explained by a broken gauge
symmetry.

We address this question in the following by investigating the
interference properties of two independent Bose condensates. It will
be shown, that the notion of a macroscopic wavefunction gives the
proper intuitive understanding about a single run of such a fictitious
experiment, no matter whether Eq.\ (\ref{rhoBECcoh}) or Eq.\
(\ref{rhoBEC}) is assumed. This indicates that a degeneracy of the
ground state should not be essential for the explanation of the
phenomena, that are usually regarded as an evidence for a
spontaneously broken gauge symmetry.

\subsection{Interference of two plane waves}
We assume that two independent atomic beams, being Bose condensates,
merge on a planar detection screen. The accumulation of a large number
of particles on the screen during a given detection time $\tau_d$ will
be regarded as a {\em single} run of the experiment. Only if we repeat
the same experiment several times with a complete physical reset
between different counting intervals, this will be called an {\em
ensemble}. Since we assume a homogeneous velocity distribution of the
incoming atoms, the distribution of particles, accumulated on the
screen during $\tau_d$ in a {\em single} run is proportional to the
line density $N(x)$ along the $x$-axis.

We describe the condensates by two independent macroscopically
occupied momentum modes with $x$-momenta $\pm k_0$ and average
occupation numbers $N_i=\left<a^\dagger_ia_i\right> (i=1,2)$. Their
number fluctuations are assumed to be negligible in comparison with
the mean occupation numbers:
\begin{equation}
\label{numfluct}
\frac{\left<a^\dagger_ia^\dagger_ia_ia_i\right>}{N_iN_i}=1+
O\left(\frac{1}{N_i}\right).
\end{equation}
This includes a representation of the condensates
in terms of Eq.\ (\ref{rhoBECcoh}) or alternatively of Eq.\
(\ref{rhoBEC}).

The first step of our analysis is to determine the {\em ensemble}
behaviour of the system. It is easily seen from
\begin{equation}
\label{ensdens}
\left<N(x)\right>V^{1/3}=N_1+N_2
\end{equation}
that {\em no} interference is visible if the measurements of many runs
are summed up. This result is no surprise due to the translational
invariance of the system in $x$-direction. Usually, also a single run
is sufficiently characterized by $\left<N(x)\right>$. However, our
considerations about broken gauge symmetry have already indicated that
this may be wrong in the case of BEC. This implies that the density
distribution $N(x)$, measured in a single run, might show a
nontrivial spatial dependence. The interesting question is therefore,
whether interference exists in the counts from a {\em single} run of
the experiment. A rigorous answer can be given by analyzing
correlation functions or conditional probabilites like
$\left<N(x)N(x+\Delta)\right>$. If these show a nontrivial dependence
on the relative distance $\Delta$, it is clear that the individual
density distribution $N(x)$ must be different from the ensemble
average Eq.\ (\ref{ensdens}). However, before turning to the detailed
discussion of correlation functions, we describe briefly the
predictions of the broken gauge symmetry model for our situation.

\subsubsection{Intuitive treatment}
Here, we assume the notion of a broken gauge symmetry to be
true. As a consequence, the macroscopic wave function of a single
condensate would be a classical plane wave with an arbitrary but fixed
phase factor. If we describe the full state vector by a product of two
coherent states, corresponding to the two condensates, we get the
combined macroscopic wave function
\begin{equation}
\label{sumwf}
\psi(x)\sqrt{V^{1/3}} = e^{i\alpha_1}\sqrt{N_1}e^{ik_0x}+
e^{i\alpha_2}\sqrt{N_2}e^{-ik_0x},
\end{equation}
which is a sum of the single macroscopic wave functions.
It is obvious that this wave function leads to an interference pattern
\begin{equation}
\label{Ndistrib}
N(x)V^{1/3}=\left(N_1+N_2\right)+2\sqrt{N_1N_2}
\cos(2k_0x+\varphi)
\end{equation}
on the screen which is determined by the {\em relative} phase
$\varphi=\alpha_1-\alpha_2$. This relative phase varies between
different runs and thus leads to Eq.\ (\ref{ensdens}) in the ensemble.
The unpredictable variation of $\varphi$ between different runs would
be usually interpreted as an experimental evidence for a broken gauge
symmetry.

However, the intuitive treatment using Eq.\ (\ref{sumwf}) depends on
assumptions, that were already questioned above. First, a degeneracy
of the ground state is assumed, so as to decompose the density matrix
into coherent states (cf.\ Eq.\ (\ref{rhoBECcoh})). Then the single
condensates are identified with specific {\em coherent} states with
the intention that their expectation value $\left<\Psi^\dagger
(x)\Psi(x)\right>_S$ characterizes the density distribution of a
single run. Regrettably, this procedure is not in rigorous agreement
with atom number conservation. These demand that the
condensate of a single run has a definite atom number, although we may
not know it. This number is allowed to fluctuate only between
different runs of the experiment. Taking number conservation serious,
one might be misled to suggest that the density distribution of a
single run is properly characterized by the expectation value
$\left<\Psi^\dagger (x)\Psi(x)\right>_S$ of {\em Fock} states. This
would yield
\begin{equation}
\label{Nfock}
N(x)V^{1/3}=N_1+N_2
\end{equation}
even for a single run, which is in contradiction with Eq.\
(\ref{Ndistrib}).

It will be the result of the rest of this section to confirm the
implications of Eq.\ (\ref{sumwf}) rigorously without using any of the
mentioned assumptions. A counterintuitive implication of this
derivation is that Eq.\ (\ref{Ndistrib}) instead of Eq.\ (\ref{Nfock})
gives the proper understanding of the single system behaviour even if
the two condensates are described by Fock states. In addition, the
following analysis illustrates, how one can extract detailed
information about the behaviour of {\em single} systems from the
quantum mechanical ensemble description.

\subsubsection{Correlation function analysis}
\label{singex}
Our aim, to characterize single runs of an experiment, confronts us
with the problem that quantum mechanics by definition determines only
properties of the ensemble, i.e.\ averages over many runs.
Nevertheless, ensemble averages do allow to predict certain features
of single runs which are not apparent in the ensemble by the analysis
of correlation functions. Such an approach has found widespread use in
quantum optics \cite{gardiner91}. Characteristic features of the time
evolution of a single atom could be derived in the case of the quantum
jumps \cite{schenzle86a,carmichael89}, eventually leading to the
numerical tool of the Monte-Carlo wavefunction technique
\cite{molmer92a,zoller92a,carmichael93}. Here, we consider the
opposite limit where even a single run of the experiment involves a
large number of particles. This implies that a complete spatial
interference pattern can be built up in a single run. In our special
situation we will be able to give a full analytical characterization
of these interference patterns. An alternative approach would be to
interprete the correlation functions in terms of conditional
probabilities. Numerical simulations of single atom counts, based on
these conditional probabilities, would lead to a similar result.

In accordance with the experimental experience, we describe the
density distribution $N(x)$, measured in a single run of the
experiment, as a smooth pattern $\tilde{N}(x)$, blurred by some
additional unpredictable and uncorrelated shot noise $\xi(x)$
\begin{equation}
\label{Nsep}
N(x)=\tilde{N}(x)+\xi(x).
\end{equation}
The shot noise accounts for the discreteness of the particles and is
assumed to obey Poissonian statistics with
\begin{eqnarray}
\Big<\xi(x)\Big>&=&0\nonumber\\
\label{noise}
\Big<\xi(x)\xi(x+\Delta)\Big>&=&\Big<N(x)\Big>\delta(\Delta).
\end{eqnarray}
Note, that $N(x)$ is a measured quantity and not a quantum mechanical
operator or an ensemble average. This implies that both $\tilde{N}(x)$
and $\xi(x)$ may differ between individual runs of the experiment.
Since the individual noisy contribution $\xi(x)$ is unpredictable, all
we can hope to achieve, is a complete characterization of the smooth
patterns $\tilde{N}(x)$.

It was indicated above, that an analysis of correlation functions
will be required to make substantial statements about the behaviour
of single runs. The lowest order correlation function is given by
\begin{equation}
\label{corrNorg} \mbox{corr}N(x,\Delta)=\Big<N(x)N(x+\Delta)\Big>,
\end{equation}
where the angular brackets denote an average over several runs of the
experiment. We can calculate this correlation by identifying the
measured density distribution $N(x)$ with the corresponding quantum
mechanical operator $\Psi^\dagger(x)\Psi(x)$. This yields
\begin{equation}
\label{corrNorgqm}\mbox{corr}N(x,\Delta)=
\Big<\Psi^\dagger(x)\Psi(x)\Psi^\dagger(x+\Delta)\Psi(x+\Delta)\Big>.
\end{equation}
Since we aim to characterize the smooth patterns $\tilde{N}(x)$, we
are interested in correlations of $\tilde{N}(x)$, defined analogously
to Eq.\ (\ref{corrNorg}), instead of those of the measured density
$N(x)$. Therefore, we rewrite Eq.\ (\ref{corrNorg}) using Eq.\
(\ref{Nsep}) and Eq.\ (\ref{noise}) as
\begin{eqnarray}
\label{corrN}
\mbox{corr}N(x,\Delta) &=& \mbox{corr}\tilde{N}(x,\Delta) +
\Big<\xi(x) \xi(x+\Delta)\Big>\\
&=&\mbox{corr}\tilde{N}(x,\Delta) +
\Big<N(x)\Big>\delta(\Delta).\nonumber
\end{eqnarray}
Normal ordering of the field operators in Eq.\ (\ref{corrNorgqm})
generates a term that is identical with
the delta correlated noise in Eq.\ (\ref{corrN}). The correlation
function for the {\em smooth} pattern is thus given by
\begin{equation}
\label{corrNN}\mbox{corr}\tilde{N}(x,\Delta) =
\Big<\Psi^\dagger(x)\Psi^\dagger(x+\Delta)\Psi(x+\Delta)\Psi(x)\Big>.
\end{equation}
We will use this equation in the following to determine the
distribution of all smooth patterns $\tilde{N}(x)$ that can be seen in
single runs of the experiment.

Since the condensates are described as macroscopically occupied
momentum modes, we have to link the above description in position
space with one in momentum space. The momentum decomposition of the
quantum mechanical field operator along the $x$-axis on the screen is
\begin{equation}
\Psi(x)=\frac{1}{\sqrt{V^{1/3}}}\sum_ke^{ikx}a_k
\end{equation}
while any possible smooth interference pattern can be
described by the Fourier decomposition
\begin{equation}
\label{NFourier}
\tilde{N}(x)=\frac{1}{V^{1/3}}\sum_ke^{ikx}\tilde{n}_k.
\end{equation}
Note that $\tilde{n}_k$ is a complex random variable with yet unknown
statistics. It satisfies the relation $\tilde{n}_k=\tilde{n}_{-k}^*$
to ensure that $\tilde{N}(x)$ is real. In the following we will
derive the statistics of the Fourier components $\tilde{n}_k$
by a comparison of the higher moments of $\tilde{n}_k$ with
higher moments of the quantum mechanical operators $a_k$.

Identification of the average over the measured screen patterns
$\left<\tilde{N}(x)\right>$ with the quantum mechanical prediction
$\left<\Psi(x)^\dagger\Psi(x)\right>$
\begin{equation}
\frac{1}{V^{1/3}}\sum_ke^{ikx}\left<\tilde{n}_k\right>=
\frac{1}{V^{1/3}}\sum_ke^{ikx}\sum_{k^\prime}\left<a^\dagger_{k^\prime
-k}a_{k^\prime}\right>
\end{equation}
yields
\begin{equation}
\left<\tilde{n}_k\right>=\sum_{k^\prime}\left<a^\dagger_{k^\prime
-k}a_{k^\prime}\right>
\end{equation}
due to the uniqueness of the Fourier decomposition. Since the two
momentum modes are assumed to be independent we get
\begin{equation}
\label{avnk}
\left<\tilde{n}_k\right>=\delta_{k,0}\left(N_1+N_2\right).
\end{equation}
Note, that the {\em ensemble} averages $\left<a_k\right>$ vanish. In a
similar way the correlation of the measured patterns corr$\tilde{N}$
\begin{equation}
\label{corrN2cl}
\mbox{corr}\tilde{N}(x,\Delta)V^{2/3}=\sum_{kk^\prime}e^{i(k+k^\prime)x}
e^{ik\Delta}
\left<\tilde{n}_k\tilde{n}_{k^\prime}\right>
\end{equation}
can be identified with its quantum mechanical analogon
\begin{eqnarray}
\label{corrN2}
\mbox{corr}\tilde{N}(x,\Delta)V^{2/3}&=&
\left<a^\dagger_1a^\dagger_1a_1a_1\right> +
\left<a^\dagger_2a^\dagger_2a_2a_2\right>\\
&&+2N_1N_2\left(1+\cos 2k_0\Delta\right).\nonumber
\end{eqnarray}
Eq.\ (\ref{corrN2}) states that the probability
for observing an atom at a position $x+\Delta$ after having observed a
previous atom at position $x$ is a periodic function of $2k_0\Delta$.
Already this observation implies that the density distribution of a
single run must show some spatial variation.
Eq.\ (\ref{corrN2cl}) and Eq.\ (\ref{corrN2}) determine the only
nonvanishing coefficients
$\left<\tilde{n}_k\tilde{n}_{k^\prime}\right>$
\begin{eqnarray}
\left<\tilde{n}_0^2\right>&=&\left<a^\dagger_1a^\dagger_1a_1a_1
\right> + \left<a^\dagger_2a^\dagger_2a_2a_2\right> + 2N_1N_2,\\
\label{a2k0}
\left<|\tilde{n}_{\pm 2k_0}|^2\right>&=&N_1N_2.
\end{eqnarray}
In the case of {\em small number fluctuations}, as assumed in Eq.\
(\ref{numfluct}), we get
\begin{equation}
\left<\tilde{n}_0^2\right>=(N_1+N_2)^2=\left<\tilde{n}_0\right>^2.
\end{equation}
This equality implies that $\tilde{n}_0$ does not fluctuate. Rather
than being a random variable, its value is invariably given by
\begin{equation}
\label{a0}
\tilde{n}_0=N_1+N_2.
\end{equation}
Eq.\ (\ref{a2k0}) and the vanishing expectation value of
$\tilde{n}_{2k_0}$ Eq.\ (\ref{avnk}) show that at least the phase of
the Fourier coefficient $\tilde{n}_{2k_0}$ is subject to fluctuations.
A similar evaluation of the fourth order correlation function yields
\begin{equation}
\left<|\tilde{n}_{2k_0}|^4\right>=(N_1N_2)^2.
\end{equation}
We skipped the explicit calculation here. One sees by a comparison
with Eq.\ (\ref{a2k0}) that the {\em modulus} of $\tilde{n}_{2k_0}$
does not fluctuate either. It is fixed to the value
\begin{equation}
\label{a2k}
|\tilde{n}_{2k_0}|=\sqrt{N_1N_2}.
\end{equation}
Since the classical correlation functions are of even order in the
atomic field operators, any phase factors cancel and no constraint for
the phase of $\tilde{n}_{2k_0}$ exists.

Thus, Eq.\ (\ref{a0}) and Eq.\ (\ref{a2k}) provide us with the
complete obtainable information about the Fourier coefficients of the
interference patterns (cf.\ Eq.\ (\ref{NFourier})) that can be
measured in single runs of the experiment. Since no constraint for the
phase of $\tilde{n}_{2k_0}$ exists, its value $\varphi$ is an equally
distributed random variable. The result of this
derivation is, that the distribution of all possible
recorded {\em smooth} interference patterns $\tilde{N}(x)$, rather
than the full noisy patterns $N(x)$, is given
by Eq. (\ref{Ndistrib}).

A surprising consequence is that with Eq.\
(\ref{Ndistrib}) holds (up to order $1/N_i$) no matter whether the
density matrices of the individual modes are given by Eq.\
(\ref{rhoBECcoh}) or Eq.\ (\ref{rhoBEC}). This shows that a Fock
state will yield in the limit of large particle numbers virtually the
same experimental results as an incoherent superposition of coherent
states (see also Ref.\ \cite{javanainen95}). The neglected terms of
order $1/N_i$ mainly account for the
existence of some correlations in the earlier eliminated noise
$\xi(x)$ in order to preserve the total particle number precisely.

We close this section with a few remarks about the range of validity
of the above result under more realistic conditions. We have assumed
in our calculations that the fluctuations in the flux are negligible
during the detection time $\tau_d$. Inclusion of these fluctuations
will clarify the range of validity of the above considerations. When
taking into account that the flux of particles may vary during the
formation of the interference pattern, the correlation function Eq.\
(\ref{corrNorg}) changes to
\begin{equation}
\label{corrNt}\mbox{corr}N(x,t;\Delta,\tau)
=\Big<N(x,t)N(x+\Delta,t+\tau)\Big>.
\end{equation}
Eq.\ (\ref{corrN2})
can then be rewritten in terms of the time correlation functions that
characterize the dynamics of the two independent modes $(i=1,2)$
\begin{eqnarray}
\mbox{corr}\tilde{N}(\Delta,\tau)V^{2/3} &=&
N_1^2g_1^{(2)}(\tau)+
N_2^2g_2^{(2)}(\tau)\nonumber\\
&&+ 2N_1N_2\left(1+\cos 2k_0\Delta Re\left\{
g_1^{(1)}(\tau)g_2^{(1)*}(\tau)\right\}\right)
\end{eqnarray}
with
\begin{eqnarray}
g_i^{(1)}(\tau)&=&\frac{\left<a^\dagger(t)a(t+\tau)
\right>}{\sqrt{\left<
a^\dagger(t)a(t)\right>\left<a^\dagger(t+\tau)a(t+\tau)\right>}}
\stackrel{\tau\to\infty}{\longrightarrow}0\nonumber\\
g_i^{(2)}(\tau)&=&\frac{\left<a^\dagger(t)a^\dagger(t+\tau)a(t+\tau)
a(t) \right>}{\left<a^\dagger(t)a(t)\right>
\left<a^\dagger(t+\tau)a(t+\tau)\right>}\stackrel{\tau\to\infty}
{\longrightarrow}1.
\end{eqnarray}
It is evident that the decay of the single interference pattern during
a longer exposure time is intimately connected to the decay of the
amplitude correlation function $g^{(1)}(\tau)$. Such a decay will
happen in any finite physical system. $g^{(2)}(\tau)$ will decay on a
similar time scale and we get
\begin{equation}
\lim_{\tau\to\infty}\mbox{corr}\tilde{N}=\left(N_1+N_2\right)^2.
\end{equation}
This implies that even a single run is appropriately described by the
quantum mechanical ensemble average
$\left<\Psi^\dagger(x)\Psi(x)\right>$, as we are used from ordinary
atomic interference experiments. However, in a system with a strong
quantum degeneracy like a laser or a Bose condensate the decay of
$g^{(1)}(\tau)$ may be so slow compared to the flux of atoms so that
intermediate time scales are relevant for experimental purposes.
The concept of broken symmetry applies only for these time scales.

\section{Interference of two localized condensates}
\label{numsim}
In the last section we approximated the condensates by two plane
waves. The general result, however, is equally valid for more
realistic setups. Here we show by a numerical simulation that the
setup of \cite{ketterle95} indeed implements such an interference
experiment, although at present no interference has been observed due
to an insufficient experimental resolution.

The experiment, reported in \cite{ketterle95}, can be summarized as
follows. Two independent Bose condensates of $^{23}$Na atoms are
stored in a magnetic trap, separated by a laser beam. These
constraints can be modelled by two identical harmonic potentials with
angular frequencies $\nu=2\pi\times 745s^{-1}$, $2\pi\times
235s^{-1}$, $2\pi\times 410s^{-1}$ in the $x, y,$ and $z$ directions,
respectively. These potentials are separated along the $x$-axis by
$x_0=100\mu$m. The total number of atoms is assumed to be
$N_0=1.5\times 10^5$, or $0.75\times10^5$ atoms in each condensate.
The scattering length, characterizing the atomic interaction, amounts
to $a=4.9$nm. After switching off the constraints, the atomic clouds
expand and eventually overlap, both due to the quantum mechanical
spreading of their wave packets and to the atomic repulsion. The final
spatial distribution of the atoms is measured after $6\mbox{ms}$ or
equivalently $28\nu_x^{-1}$.

In our calculation we simulate the expansion of these atomic clouds.
We assume that the two condensates initially are at zero temperature,
with both condensates equally populated. The investigation of atomic
interference at nonvanishing temperatures is put forward to the next
section. There, it is shown that interference can be observed in a
different setup also at small but finite temperatures.

The sodium atoms are subject to a repulsive interaction. The initial
wave function of a single condensate is therefore described by a
solution of the
Gross-Pitaevskii equation \cite{fetter71}
\begin{equation}
\label{grosspit}
\mu\psi({\vec{r}},t)=\left(-\frac{\hbar^2\vec\nabla^2}{2m}+
V(\vec{r}) +\tilde{U}|\psi( {\vec{r}},t)|^2 \right)\psi({\vec{r}},t)
\end{equation}
with the chemical potential $\mu$ and
\begin{equation}
\tilde{U}=N_0\frac{4\pi\hbar^2a}{m}.
\end{equation}
Note that the macroscopic wave function $\psi({\vec{r}},t)$ is
normalized to one in this notation. The chemical potential
\begin{equation}
\mu = 0.118\frac{\tilde{U}}{N_0}\left(\frac{N_0}{\Delta x^2\Delta
y^2\Delta z^2a^{3/2}}\right)^{2/5}
\end{equation}
is expressed in terms of the natural length units of the harmonic
potential $V({\vec{r}})$ along the coordinate axes. In the case of the
$x$-direction, for example, this length unit is
\begin{equation}
\Delta x=\sqrt{\frac{\hbar}{m\nu_x}}=0.78\mu\mbox{m}.
\end{equation}

Since the total energy is dominated by the particle interactions and
the external harmonic potential, it is a good first approximation
\cite{griffin95} to neglect the kinetic energy in Eq.\
(\ref{grosspit}). This yields the approximate expression
\begin{equation}
\label{groundapp}
\psi({\vec{r}})\approx\sqrt{\Big(\mu-V({\vec{r}})\Big)/\tilde{U}}
\end{equation}
for the initial wave function, provided that the expression under the
square root is positive. Otherwise, the wave function vanishes
abruptly, for example on the $x$-axis at a distance of $3.24\mu$m
from the origin. The physical wave function, however, vanishes
smoothly on a small scale that is called the healing
length\cite{griffin95}. This smooth rather than abrupt vanishing
accounts for the existence of a limited kinetic energy.

We therefore conclude that the initial wave function of a single
condensate is largely of a parabolic shape and about 4 times larger
than the corresponding noninteracting wave function. Note, that this
significant deviation from the pure harmonic oscillator ground state
is due to a balance between the potential energies of the atom
interactions and the external harmonic potential. It should not be
confused with the depletion of the condensate that is found in an
interacting homogenous gas \cite{fetter71}:
\begin{equation}
\frac{N-N_0}{N}=\frac{8}{3}\sqrt{\frac{n a^3}{\pi}}.
\end{equation}
Inserting the peak densities of the current experiment into this
formula indicates a depletion of less than 1\% which is negligible
for our purposes.

According to the results of the last section it is justified to
describe the two independent condensates by one combined macroscopic
wave function, for characterizing a single run of the experiment. This
wave function consists of two spatially separated parts, corresponding
to the two condensates, that are joined with a relative phase
$\varphi$ between them. Of course, this relative phase varies between
different runs of the experiment. Since variations in the relative
phase lead only to trivial shifts of the interference pattern we
restrict ourselves to the case $\varphi=0$.

The trap is switched off at a given time and the atomic clouds expand.
We describe this expansion assuming the validity of the nonlinear
Schr\"odinger equation \cite{griffin95}
\begin{equation}
\label{nlse}
i\hbar\frac{\partial}{\partial
t}\psi({\vec{r}},t)=\left(-\frac{\hbar^2\vec\nabla^2}{2m}+\tilde{U}|\psi(
{\vec{r}},t)|^2 \right)\psi({\vec{r}},t).
\end{equation}
This is also referred to as the time-dependent Gross-Pitaevskii
equation (cf.\ Eq.\ (\ref{grosspit})). Its formal structure resembles
a mean-field equation when replacing the original two-body potential
by a pseudopotential. It can be derived for cold and dilute gases by
applying the ladder approximation to the two-body Green's function
\cite{fetter71}. Even within this approximation Eq.\ (\ref{nlse}) does
not account for the full dynamics, but it should be a good
approximation for time scales, where collisional rates can be
neglected.\footnote{Note, that the atomic clouds are not in thermal
equilibrium during the described expansion. The earlier assumption of
zero temperature applies only in the presence of the trapping
potentials.}

Due to the nonlinearity in Eq.\ (\ref{nlse}), it is incorrect
to separate the wave function with respect to the spatial
coordinates. Nevertheless, a separation of the spatial degrees of
freedom has to be applied as an approximation, since a full
three-dimensional numerical treatment of the problem is far beyond
reach. Such an approximation goes back to \cite{dirac30} and has
found widespread use in chemical physics where it is called
time-dependent self-consistent field (TDSCF) method. Its validity has
been studied for example in \cite{alimi90}. This approximation
reduces the propagation of a three-dimensional wave function to that
of three coupled one-dimensional wave functions. The error of this
simplification depends on the deviation of the wave function from a
Gaussian distribution.

The numerical propagation of the wave function is performed by a split
operator technique \cite{pathria90} that is accurate and easy to
implement. The initial wave function is generated by an imaginary time
propagation of Eq.\ (\ref{nlse}), which yields the full solution of
Eq.\ (\ref{grosspit}) in contrast to the approximate wave function
Eq.\ (\ref{groundapp}).

We plot in Fig.\ \ref{Xdistft} the time evolution of the $x$
distribution, as it could have be seen in the experiment. We have
averaged over the relative phase of the two condensates since the
fringes are not resolvable on the plotted scale. The two wave
functions overlap after about 3ms. Even though there is a considerable
overlap of the two condensates, it is not large enough to allow for an
interpretation of the $x$ distribution in terms of a velocity
distribution of the initial single condensates.

In Fig.\ \ref{Xdist} we show both the initial and final $x$
distribution including the interference fringes as they are predicted
by the nonlinear Schr\"odinger equation. The periodic length of the
fringes is about 1$\mu$m and thus below the pixel resolution of
12.4$\mu$m used in the experiment of \cite{ketterle95}. The period of
the fringes
increases slowly with time. However, its size is of the same order of
magnitude as the healing length throughout the whole investigated
time interval.
The period of the fringes could be enlarged by widening the trap
potentials before dropping the atoms. This would reduce both the
initial kinetic and interaction energy.

Fig.\ \ref{XZcalc} shows the calculated $x$-$z$ distribution after
6ms. Its square like shape is an artifact of the TDSCF
approximation. A full three-dimensional calculation would yield a
slightly
larger extension along the axes. For comparison,
we plot in Fig.\ \ref{XZexp} the corresponding distribution,
calculated from the experimental data of \cite{ketterle95}. The
original data describe the spatial dependence of the light absorption.
Assuming an exponential absorption law we have transformed the light
absorption into a particle distribution and normalized it for a better
comparison with Fig.\ \ref{XZcalc}. We conclude that the experimental
data of \cite{ketterle95} are in good agreement with a description
using the nonlinear Schr\"odinger equation. While the existence of
shoulders and their positions are in especially good agreement, also
the total size of the atomic cloud is fairly well reproduced. A
detailed quantitative analysis, however, requires a better
accuracy of the experimental data.

\section{Interference properties of a single Bose condensate}
\label{visib}

In this Section we analyze the effects of finite temperatures in
interference experiments with a Bose gas.  We consider a sample of $N$
bosons confined in a three-dimensional harmonic trap, not necessarily
isotropic. After reaching thermal equilibrium at a given temperature
$T$, the gas is released and the sample falls through two pinholes,
separated by a distance $2d$, that are located on a screen $S_1$ at a
distance $L$ from the trap center (Fig.\ \ref{figsetup}). The atoms
are recorded on a second screen $S_2$, at a distance $D$ from the trap
center. We wish to stress the fact that in this experiment the
wavefunctions coming from both pinholes have fixed relative phases
since they represent filtered parts from the same large wavefunction
that was stored in the trap. Therefore, different realizations of the
experiment will show the same interference pattern. This is in
contrast with the experiments analyzed in Sec.\ \ref{twocond}, in
which we considered two {\it independent} condensates. We will not
perform the correlation analysis of Sec.\ \ref{twocond} here since it
does not yield additional information in this case. The main purpose
of this Section is to show how the interference pattern depends on the
temperature of the sample, and to what extent it reflects the phase
transition at the critical temperature. We will consider the case of
an ideal gas, i.e. we will not take into account the role of atomic
collisions. The problem of a (weakly) interacting Bose gas at finite
temperatures deserves a separate analysis.

\subsection{Evolution through the pinholes}

We are interested in the mean number of particles $dN$ per unit time
and unit area deposited at any point $\vec{r}=(x,y,D)$ on the screen
$S_2$ at a given time $t$. This quantity is given by the expectation
value of the $z$ component of the probability current,
\begin{equation}
\label{current}
I(\vec{r},t) \equiv \frac{dN}{dS \; dt} =  \frac{\hbar}{M}
{\rm Im} \langle \Psi(\vec{r},t)^\dagger
\frac{\partial}{\partial z} \Psi(\vec{r},t) \rangle,
\end{equation}
where $M$ is the atomic mass, and $\Psi(\vec{r},t)$ is the field
operator describing the sample. For the case of non--interacting
particles, this operator can always be written as
\begin{equation}
\label{HPSI}
\Psi(\vec{r},t) = \sum_{\vec{n}} \hat{a}_{\vec{n}}
\psi_{\vec{n}}(\vec{r},t),
\end{equation}
where the wavefunction $\psi_{\vec{n}}(\vec{r},t)$ satisfies the
(single particle) Schr\"odinger equation describing the evolution of a
single particle. Here, $\psi_{\vec{n}}(\vec{r},0)$ is the
eigenfunction of the three-dimensional harmonic oscillator with
quantum numbers $\vec{n}=(n_x,n_y,n_z)$ ($n=0,1,\ldots$). As usual,
$a_{\vec{n}}$ are annihilation opeartors of particles in the state
$\vec{n}$.

It is worth mentioning that, if the wavefunction is expanded in terms
of momentum eigenstates as in the previous Section, $I(\vec{r},t)$ can
be fully characterized by the density of particles. Here, however, we
find it more convenient to use the eigenstates of the harmonic
oscillator. Therefore the probability current has to be evaluated
explicitly for describing the interference pattern.

Substituting (\ref{HPSI}) into (\ref{current}), we obtain
\begin{equation}
\label{IRRtau}
I(\vec{r},t) = \sum_{\vec{n}} N_{\vec{n}} I_{\vec{n}}(\vec{r},t)
\end{equation}
implying that the incoherently populated trap eigenfunctions propagate
independently after releasing the trap. Their occupation numbers
$N_{\vec{n}}=\langle a_{\vec{n}}^\dagger a_{\vec{n}} \rangle$ are
given by the Bose--Einstein distribution
\begin{equation}
\label{BED}
N_{\vec{n}} = \frac{\lambda e^{-\beta \hbar (n_x \nu_x +n_y \nu_y
+n_z
\nu_z)}} {1-\lambda e^{-\beta \hbar (n_x \nu_x +n_y \nu_y +n_z
\nu_z)}}
\end{equation}
where $\beta=1/\kappa_B T$, $\lambda$ is related to the number
of particles in the trap $N$, and $\nu_{x,y,z}$ the trap frequencies
along the three axes (see Appendix B). We have also defined
\begin{equation}
\label{INN}
I_{\vec{n}}(\vec{r},t) = \frac{\hbar}{M}  {\rm Im}
\left[\psi_{\vec{n}}(\vec{r},t)^\ast
\frac{\partial}{\partial z} \psi_{\vec{n}}(\vec{r},t)\right].
\end{equation}
In view of this formula, we only need to derive the expression for
$\psi_{\vec{n}}(\vec{r},t)$; that is, we simply have to find the
evolution through the pinholes of a single particle that is initially
in the harmonic oscillator eigenstate with quantum numbers $\vec{n}$.

Apart from neglecting the role of collisions, in order to calculate
the evolution of $\psi_{\vec{n}}(\vec{r},0)$ we will make other
simplifying assumptions. First, we will assume that $L$ is small
enough such that during the time that the sample needs to reach the
pinholes, this wavefunction basically does not change. Second, we will
assume that the size of the pinholes is small compared to the typical
distances over which this wavefunction varies. Finally, we will take
$D$ to be large enough in such a way that the parts of the
wavefunction coming from each pinhole overlaps with each other.
According to these assumptions, we can approximate
\begin{equation}
\psi_{\vec{n}}(\vec{r},0) \simeq k \psi_{n_x}(x)
[ \delta(x-x_1) + \delta(x-x_2)]
\psi_{n_y}(y) \delta(y-y_0)
\psi_{n_z}(z),
\end{equation}
where $(x_1,y_0)$ and $(x_2,y_0)$ are the coordinates of the pinholes
in the first screen $S_1$, $\psi_n(x)$ is the eigenstate of the
one-dimensional harmonic oscillator in position representation, and
$k$ a normalization constant.

After a time $t>0$, the wavefunction evolves to
\begin{equation}
\label{PsiNNt}
\psi_{\vec{n}}(\vec{r},t) \simeq k
[ G_0(x,x_1,t) \psi_{n_x}(x_1) + G_0(x,x_2,t) \psi_{n_x}(x_2)]
G_0(y,y_0,t)\psi_{n_y}(y_0)
\psi_{n_z}(z,t),
\end{equation}
where
\begin{equation}
\label{psiz}
\psi_{n_z}(z,t)= \int_{-\infty}^\infty dz' G_g(z,z',t)
\psi_{n_z}(z').
\end{equation}
Here,
\begin{mathletters}
\begin{eqnarray}
\label{G0}
G_0(x,x',t) &=& e^{-i\pi/4}
\left[ \frac{M}{2\pi\hbar t} \right]^{1/2}
e^{i \frac{M(x-x')^2}{2\hbar t}} \\
G_g(z,z',t) &=& e^{-i\pi/4}
\left[ \frac{M}{2\pi\hbar t} \right]^{1/2}
e^{i \frac{M}{2\hbar t} [(z-z')^2+gt^2(z+z')-g^2t^3/12]}
\end{eqnarray}
\end{mathletters}
are the free propagator and the propagator under a constant
force, respectively. The substitution of (\ref{PsiNNt}) into
(\ref{INN}) allows us to write
\begin{equation}
I_{\vec{n}}(\vec{r},t) = k^2 I_{n_x}^x(x,t) I_{n_y}^y(y,t)
I_{n_z}^z(z,t)
\end{equation}
where
\begin{mathletters}
\begin{eqnarray}
I_{n_x}^x(x,t) &=&
|G_0(x,x_1,t) \psi_{n_x}(x_1) + G_0(x,x_2,t) \psi_{n_x}(x_2)|^2,\\
I_{n_y}^y(y,t) &=& |G_0(y,y_1,t)\psi_{n_y}(y_0)|^2, \\
\label{Inz}
I_{n_z}^z(z,t) &=&  \frac{\hbar}{M} {\rm Im} \left[ \psi_{n_z}
(z,t)^\ast \frac{\partial}{\partial z}\psi_{n_z}(z,t) \right].
\end{eqnarray}
\end{mathletters}

Let us now derive simple expressions for these quantities. First,
using (\ref{G0}) we find
\begin{equation}
I_{n_y}^y(y,t)= (M/2\pi\hbar t)|\psi_{n_y}(y_0)|^2
\end{equation}
for all $y$ \cite{note}.
On the other hand, as it is shown in the Appendix A,
\begin{equation}
\label{Inz2}
I_{n_z}^z(z,t)= F(t) |\tilde\psi_{n_z}(z-gt^2/2,0)|^2,
\end{equation}
where
\begin{equation}
\label{Ft}
F(t)= \frac{gt}{1+(\nu_z t)^2}
\left( 1 + \frac{\nu_z^2}{g}\left[ \frac{1}{2} g t^2
- z \right]\right),
\end{equation}
and $\tilde\psi_{n_z}$ is the eigenstate of the one-dimensional
harmonic oscillator with frequency $\nu_z/[1+(\nu_z t)^2]$. Finally,
\begin{equation}
I_{n_x}^x(x,t) = (M/2\pi\hbar t) \{ |\psi_{n_x}(x_1)|^2 +
|\psi_{n_x}(x_2)|^2 + 2 \psi_{n_x}(x_1) \psi_{n_x}(x_2)
\cos [ 2\pi x/x_f(t)+ \phi(t) ] \}
\end{equation}
where we have taken into account that $\psi_{n_x}(x)$ is real
\cite{note}. Here,
\begin{equation}
\label{xF}
x_f(t) = \frac{\pi \hbar t}{Md} = 2\pi \nu_x t
\frac{a_{0x}^2}{d}
\end{equation}
with $a_{0x}=[\hbar/(2M\nu_x)]^{1/2}$ being the size of the ground
state wavefunction, and $\phi(t)=M(x_1^2-x_2^2)/(2\hbar t)$.

\subsection{Interference and visibility}

Let us now specialize the above derived expressions to a simple case.
We take $y_0=0$ and $x_1=d=-x_2$ (i.e. the pinholes are symmetrically
situated along the $x$ axis). For the sake of simplicity we choose the
time $\tau=(2D/g)^{1/2}$, which is the time required for the center of
the trap to reach by free fall the screen $S_2$. In this case, we
obtain
\begin{mathletters}
\begin{eqnarray}
\label{Ixx}
I_{n_x}^x(x,\tau) &\propto& |\psi_{n_x}(d)|^2
\{1 + (-1)^{n_x} \cos[2\pi x/x_f(\tau)] \},\\
I_{n_y}^y(y,\tau) &\propto& |\psi_{n_y}(0)|^2, \\
I_{n_z}^z(z,\tau) &\propto& |\psi_{n_z}(0)|^2,
\end{eqnarray}
\end{mathletters}
Note that the term $(-1)^{n_x}$ appearing in the first of these
expresions comes from the fact that $\psi_{n_x}(-d)= (-1)^{n_x}
\psi_{n_x}(d)$. Thus, depending on $n_x$ being even or odd, the
interference pattern has a maximum or a minimum at $x=0$. It is
obvious that the incoherent addition of all the contributions for
different quantum numbers $n_x$ [see (\ref{IRRtau})] will decrease the
visibility. Therefore, the origin of the temperature dependence of the
interference fringes is precisely the term $(-1)^{n_x}$ accompanying
the cosine. For sufficiently low temperature, the Bose--Einstein
distribution (\ref{BED}) is peaked at $n_x=0$, and therefore fringes
along the $y$ axis will be clearly displayed on the screen $S_2$. The
separation of the fringes is $x_f(\tau)$ given in (\ref{xF}). Note
that, in principle, this separation changes with time. In practice,
this variation will be negligible as long as the time required by the
sample to cross the screen $\delta \tau$ is much smaller than $\tau$.
More specifically, we can neglect this time dependence for
\begin{equation}
\frac{\delta\tau}{\tau}=\sqrt{n_z[1+(\nu\tau)^2]}
\frac{a_{0z}}{2D} \ll 1.
\end{equation}
This condition can always be fulfilled for a sufficiently long
distance $D$ (note that $\tau$ scales as $\sqrt{D}$, and therefore
$\delta\tau/\tau$ scales as $1/\sqrt{D}$).

Let us now analyze the temperature dependence of the visibility
\begin{equation}
V= \frac{I^{\rm max}-I^{\rm min}}{I^{\rm max}+I^{\rm min}}.
\end{equation}
where $I$ is given in (\ref{IRRtau}). The maxima and minima of the
expression (\ref{Ixx}) can be easily calculated, taking into account
that for fixed $n_{y,z}$, $N_{\vec{n}}$ is a decreasing function of
$n_x$. Thus, the maxima (minima) in the fringes correspond to
$x_k^{\rm max}= k x_f(\tau)$ [$x_k^{\rm min} = (k+1/2) x_f(\tau)$],
with $k=0,\pm 1,\ldots$. Using this result we find that
\begin{mathletters}
\label{Imaxmin}
\begin{eqnarray}
I^{\rm max}+I^{\rm min} &\propto&  \sum_{\vec{n}} N_{\vec{n}}
|\psi_{n_x}(d)|^2 |\psi_{n_y}(0)|^2  |\psi_{n_z}(0)|^2,\\
I^{\rm max}-I^{\rm min} &\propto&  \sum_{\vec{n}} (-1)^{n_x}
N_{\vec{n}} |\psi_{n_x}(d)|^2 |\psi_{n_y}(0)|^2  |\psi_{n_z}(0)|^2.
\end{eqnarray}
\end{mathletters}

In Fig.\ \ref{visplot} we have plotted the visibility (solid--lines)
as a function of the scaled temperature $\kappa_B T/(\hbar \nu)$ for
$N=10000$ particles in an isotropic harmonic oscillator and different
distances $2d$ between the two pinholes. We have also plotted
(dotted--line) the proportion of particles in the condensate $N_{0}$
(for the numerical method used to calculate these results see Appendix
B). Note first that for a given temperature, the visibility decreases
as $d$ increases. The reason for that is as follows. As mentioned
above, the eigenstates of the harmonic oscillator with even quantum
numbers $n_x$ give {\it in phase} contributions to the fringe pattern,
whereas the ones with odd $n_x$ give {\it out of phase} contributions,
i.e. tend to decrease the visibility. The effect of the terms with odd
$n_x$ tends to zero for $d\rightarrow 0$, since $\psi_{n_x}(d)$ tends
also to zero (note that this wavefunction is antisymmetric). Thus, the
pinholes select the particles whose wavefunctions have even $n_x$. As
$d$ increases, up to the order of the ground state wavefunction
$a_{0x}$, the visibility decreases since the contributions of the odd
wavefunctions becomes more important. For $d\gg a_{0x}$ the visibility
remains small even below the critical point, since only the
wavefunctions with high $n_x$ contribute. On the other hand, the
transition point of BEC is reflected in the visibility, which shows
that double--slit experiments can be used to observe the phase
transition. Note the dramatic changes at the transition point for
moderate values of $d$. It is somehow surprising that even for $N-N_0
> N_0$ the visibility already becomes very large, since one would
expect that the particles that are out of the condensate would produce
some noise that would dominate over the effect of the condensed
particles. However, this is not the case. The $N-N_0$ particles out of
the condensate are distributed among several states with different
$n_x$ (this distribution is similar to a Boltzmann distribution and
therefore has a long tail). As $n_x$ increases, their contribution to
the intensity becomes smaller and smaller given that
$|\psi_{n_x}(d)|^2 \rightarrow 0$. Thus, for $d\alt a_{0x}$ the change
of visibility at the transition point becomes very pronounced.

\section{Conclusion}
\label{conclusion}
We have shown that two independent Bose condensates, when overlapping
on a screen, can exhibit an interference pattern in a single run of
the experiment. A rigorous analysis in terms of correlation functions
confirmed that the description of an individual Bose condensate by a
coherent state, i.e.\ a macroscopic wave function, gives the proper
characterization of single runs of this interference experiment. This
applies although atomic coherent states violate fundamental
conservation laws and holds in the limit of large occupation numbers
even for a single Fock state. The macroscopic wave function concept
resembles the way Quantum Electrodynamics converges to Classical
Electrodynamics for large photon numbers. However, one fundamental
difference between atomic and photonic coherent states should be
mentioned. It is, at least in principle, possible to measure the
absolute phase of a photon coherent state, since the electric field
amplitude is a true observable. As a consequence, the photon number is
no conserved quantity. In contrast, the atom number is strictly
conserved. In return, the macroscopic wave function itself is not an
observable, thus inhibiting the measurement of its {\em absolute}
phase. Consequently, the investigated interference experiment
yielded only information about the {\em relative} phase between two
macroscopic wave functions.

It is another aim of the paper to describe simple experiments
that allow for a comparison between experimental and
theoretical features of a Bose condensate. The
considerations above, concerning the interference of independent Bose
condensates allowed the numerical simulation of a recent experiment
\cite{ketterle95} where two
independent Bose condensates are released from a magnetic trap. They
expand and eventually overlap. In our calculations we accounted for
particle interactions but assumed an initial distribution with zero
temperature. The calculations
predict interference fringes with a period below the resolution of the
current experimental data. However, the size of the final
atomic cloud is in good agreement with the experimental result.
A further quantitative analysis is limited by the present accuracy
of the experiment. By improved detection techniques it should be
possible to study the existence of the interference fringes.
This could yield substantial and new
information about the condensate and the range of validity of the
nonlinear Schr\"odinger equation.

As a possible future experiment we proposed the study of the
interference properties of a Bose condensed system in a conventional
Young interference experiment. We assumed that the atoms are released
from a harmonic trap and fall freely through two pinholes. Here we
neglected particle interactions since the above calculations have
shown that the visibility of the interference fringes is not severely
affected by these interactions. Instead we studied the dependence of
the visibility on the temperature. The transition to a Bose condensed
phase is reflected by a sharp rise in the visibility of the fringes.
Under favourable conditions the visibility reaches values close to one
immediately below the critical temperature. In addition a variation of
the distance of the the pinholes shows that the coherence length of a
finite Bose condensed system is comparable with the size of the
groundstate wavefunction.

\acknowledgments
We are very grateful to W.~Ketterle and his
coworkers for providing us with the original experimental data for a
comparison with the calculations. H.~W. and M.~N. thank T.~W.~H\"ansch
and K.~Pachucki for raising the question about the interference of
Fock states. This work was supported by the Deutsche
Forschungsgemeinschaft. J.~C. and P.~Z. were supported in part by the
Austrian Science Foundation.

\appendix

\section{Evolution in the $z$ direction}

In this appendix we derive some of the formulas used in Section
III. First, starting from (\ref{psiz}) we write
\begin{equation}
\psi_{n}(z,t) = A(z,t) K_{n} \int_{-\infty}^\infty dz'
H_n(\alpha z') e^{-R(t)z'^2+i C(z,t) z'}.
\end{equation}
Here, we have used
\begin{equation}
\psi_n(z,0)=K_n H_n(\alpha z) e^{- \alpha^2 z^2/2},
\end{equation}
where $\alpha=(M\nu_z/\hbar)^{1/2}$,
$K_n=[\alpha/(2^n n!\sqrt{\pi})]^{1/2}$, and $H_n$ is the
$n$--th Hermite polynomial. We have also defined
\begin{mathletters}
\begin{eqnarray}
A(z,t) &=& e^{-i\pi/4} \sqrt{\frac{M}{2\pi\hbar t}}
e^{iM/(2\hbar t)(z^2+gt^2z-g^2t^4/12)},\\
C(z,t) &=& \frac{M}{\hbar t} \left[\frac{1}{2} g t^2 -
z\right],\\
R(t) &=& \frac{\alpha^2}{2}-i \frac{M}{2\hbar t} =
\frac{M}{2\hbar t} (\nu_z t - i).
\end{eqnarray}
\end{mathletters}
Performing the change of variables $\sqrt{R(t)}z=x$ and
transforming the path for the integration in the complex plane,
we arrive at
\begin{equation}
\psi_{n}(z,t) = \frac{A(z,t)}{\sqrt{R(t)}} K_{n} e^{-B(z,t)^2}
\int_{-\infty}^\infty dz'
H_n\left[\frac{\alpha z'}{\sqrt{R(t)}}\right] e^{-[x-iB(z,t)]^2},
\end{equation}
where $B(z,t)=C(z,t)/[2\sqrt{R(t)}]$. Performing the
integration, and after some lenghty algebra we obtain
\begin{equation}
\label{psizt}
\psi_{n}(z,t) = \frac{A(z,t)}{\sqrt{R(t)}} K_{n}
e^{-\tilde \alpha(t)^2\tilde Z(z,t)^2/2 [1+i/(\nu_zt)]}
[-R(t)^\ast/R(t)]^{n/2} H_n [\tilde \alpha(t) \tilde Z(z,t)],
\end{equation}
where $\tilde \alpha=\alpha/\sqrt{1+\nu_z^2t^2}$ and
$Z(z,t)=z-gt^2/2$. Now, the expression (\ref{Inz2}) can be
easily derived starting from (\ref{Inz}) and using (\ref{psizt}).
To do that, one has to note that when performing the derivatives
of $\psi_{n}(z,t)$, the contribution given by the
derivative of the Hermite polynomial vanishes when taking the
imaginary part. We obtain
\begin{equation}
I_{n_z}^z(z,t)= F(t) |\psi_{n_z}(z,t)|^2,
\end{equation}
where $F(t)$ is given in (\ref{Ft}). Taking the modulus square
of (\ref{psizt}) we obtain (\ref{Inz}).

\section{Formulas for the numerical evaluation of the
visibility}

In this appendix we give some of the formulas used to evaluate
numerically the expression of the visibility for a
Bose--Einstein distribution. The total number of particles can
be written as
\begin{equation}
\label{Ntot}
N=\sum_{\vec{n}} N_{\vec{n}} = \sum_{k=1}^\infty \left[
\frac{N_0}{N_0+1}\right]^k \prod_{i=x,y,z} (1-e^{-\beta\hbar k
\nu_i})^{-1}.
\end{equation}
Given a fixed number of particles $N$ and a temperature
$T$, we have first determined the value of $N_0$ by using a
bisection method varying $N_0$ until Eq.~(\ref{Ntot}) is
verified. Thus, we can calculate the whole distribution through
Eq.~(\ref{BED}) since $\lambda=N_0/(1+N_0)$.

On the other hand, in order to calculate the expressions
(\ref{Imaxmin}) we need to determine sums of the form
\begin{mathletters}
\begin{eqnarray}
S_1&=& \sum_{n=0}^\infty e^{-\beta\hbar\nu k(n+1/2)}
|\psi_n(x)|^2,\\
S_2&=& \sum_{n=0}^\infty (-1)^n e^{-\beta\hbar\nu k(n+1/2)}
|\psi_n(x)|^2.
\end{eqnarray}
\end{mathletters}
This can be easily performed by noting that these expressions
are related to the propagator for the harmonic oscillator with
an imaginary time. In particular,
\begin{mathletters}
\begin{eqnarray}
S_1&=& G(x,x,t=-i\beta k)\\
S_2&=& G[x,x,t=\pi/(\hbar\nu)-i\beta k],
\end{eqnarray}
\end{mathletters}
where
\begin{equation}
G(x,x,t)= \sqrt{A/(2\pi)} e^{Ax^2[1-\cos(\nu t)]},
\end{equation}
with $A=M\nu/[i\hbar\sin(\nu t)]$. Using these expresions, we
find
\begin{mathletters}
\label{dldl}
\begin{eqnarray}
I^{\rm max}+I^{\rm min} &=& \sum_{k=1}^\infty
\left[ \frac{N_0}{N_0+1}\right]^k
e^{\alpha^2 d^2[1-\cosh(\beta\hbar\nu_x k)]/
\sinh(\beta\hbar\nu_x k)}
\prod_{i=x,y,z} \sqrt{\sinh(\beta\hbar\nu_i k)},\\
I^{\rm max}-I^{\rm min} &=& \sum_{k=1}^\infty
\left[ \frac{N_0}{N_0+1}\right]^k
e^{-\alpha^2 d^2[1+\cosh(\beta\hbar\nu_x k)]/
\sinh(\beta\hbar\nu_x k)}
\prod_{i=x,y,z} \sqrt{\sinh(\beta\hbar\nu_i k)},\\
\end{eqnarray}
\end{mathletters}
Note that one could use directly the formulas (\ref{Imaxmin}),
which involve 3 nested sums. However, Eq.~(\ref{dldl}) involves
a single sum, which saves a lot of computer time.

\narrowtext
\twocolumn


\begin{figure}
\caption{\label{Xdistft}Calculated time evolution of the $x$
distribution. The interference fringes are removed by averaging over
$\varphi$. The contour lines correspond to an increment of 0.001.}
\end{figure}

\begin{figure}
\caption{\label{Xdist}Initial and final $x$ distribution. The final
distribution is plotted both with a relative phase $\varphi=0$ and
averaged over all possible phases.}
\end{figure}

\begin{figure}
\caption{\label{XZcalc}Calculated final $x$-$z$ distribution with
removed interference fringes.}
\end{figure}

\begin{figure}
\caption{\label{XZexp}Experimental atomic density, calculated from the
light absorption. A renormalization was performed to facilitate
comparison with Fig.\ \protect{\ref{XZcalc}}. The original data
correspond to Fig.\ 2c of \protect{\cite{ketterle95}}. Their use is
with kind permission of W.~Ketterle and coworkers.}
\end{figure}

\begin{figure}
\caption{\label{figsetup}
Setup of the experiment considered in Section III. An ideal Bose gas
in thermal equilibrium at temperature $T$ is dropped through two
pinholes. The interference is recorded on a screen.}
\end{figure}

\begin{figure}
\caption{\label{visplot}
Visibility $V$ plotted as a function of the scaled temperature for
(curves from top to bottom) $d=0.5,1.56,2.61,3.67,4.72,5.78$, and
$6.83 [\hbar/(M\nu)]$ (the inset shows a detail). In dotted line we
have plotted the proportion of particles in the condensate
$n_0=N_0/N$.}
\end{figure}

\begin{thebibliography}{10}
\bibitem{note}
Given that we have approximated the initial state by a delta function,
this result involves a infinitely fast spreading of the wavefunction.
In reality, the results will be valid over a finite region of the
screen $S_2$ whose dimensions will increase with the distance $D$.
\end{thebibliography}
\end{document}